\documentclass{ws-ijmpa}

\usepackage{graphicx}
\usepackage{clpsref}

\newcommand{\sfrac}[2]{\mbox{\footnotesize $\displaystyle \frac{#1}{#2}$}} 
\newcommand{\lsim}{\mathrel{\rlap{\lower4pt\hbox{\hskip0pt$\sim$}} 
\raise1pt\hbox{$<$}}}           %less than or approx. symbol 
\newcommand{\gsim}{\mathrel{\rlap{\lower4pt\hbox{\hskip0pt$\sim$}} 
\raise1pt\hbox{$>$}}}           %greater than or approx. symbol 

\begin{document}

\markboth{A.~H\"oll, A.~Krassnigg, C.\,D.~Roberts and S.\,V.~Wright}
{On the complexion of pseudoscalar mesons}

\catchline{}{}{}{}{}
%
%%%%%%%%%%%%%%%%%%%%%%%%%%%%%%%%%%%%%%%%%%%%%%%%%%%%%%%%%%%%%%%%%%%%

\title{On the complexion of pseudoscalar mesons}

\author{\footnotesize A.~H\"OLL,$\!$\footnotemark[1]\,\, %
A.~KRASSNIGG,$\!$\footnotemark[1]\,\, %
C.\,D.~ROBERTS\footnotemark[1]\,\,\footnotemark[2]\,\, %
and S.\,V.~WRIGHT\footnotemark[1]%\footnote{Typeset names in 8 pt roman, uppercase. Use the footnote to indicate the present or permanent address of the author.}
}

\address{\footnotemark[1]\ Physics Division, Argonne National Laboratory,\\ Argonne IL 60439, 
USA\\
\footnotemark[2]\ Fachbereich Physik, Universit\"at Rostock,\\ D-18051 Rostock, Germany%\footnote{State completely without abbreviations, the affiliation and mailing address, including country. Typeset in 8 pt italic.}
}

%\author{SECOND AUTHOR}

%\address{Group, Laboratory, Address\\
%City, State ZIP/Zone, Country
%}

\maketitle

\pub{Received (13 November 2004)}{}

\begin{abstract}
A strongly momentum-dependent dressed-quark mass function is basic to QCD.  It is central to the appearance of a constituent-quark mass-scale and an existential prerequisite for Goldstone modes.  Dyson-Schwinger equation (DSEs) studies have long emphasised this importance, and have proved that QCD's Goldstone modes are the only pseudoscalar mesons to possess a nonzero leptonic decay constant in the chiral limit when chiral symmetry is dynamically broken, while the decay constants of their radial excitations vanish.  Such features are readily illustrated using a rainbow-ladder truncation of the DSEs.  In this connection we find (in GeV): $f_{\eta_c(1S)}= 0.233$, $m_{\eta_c(2S)}=3.42$; and support for interpreting $\eta(1295)$, $\eta(1470)$ as the first radial excitations of $\eta(548)$, $\eta^\prime(958)$, respectively, and $K(1460)$ as the first radial excitation of the kaon.  Moreover, such radial excitations have electromagnetic diameters greater than $2\,$fm.  This exceeds the spatial length of lattices used typically in contemporary lattice-QCD. 
\end{abstract}

\bigskip

%\section{Introduction}    %) A SECTION HEADING
The meson spectrum contains three pseudoscalars [$I^G (J^P) L = 1^- (0^-) S$] with masses below $2\,$GeV: $\pi(140)$; $\pi(1300)$; and $\pi(1800)$.  In the context of a model constituent-quark Hamiltonian, these mesons are often viewed as the first three members of a $Q\bar Q$ $n\, ^1\!S_0$ trajectory, where $n$ is the principal quantum number; i.e., the $\pi(140)$ is the $S$-wave ground state while the others are its first two radial excitations.  This reasoning suggests that the properties of the $\pi(1300)$ and $\pi(1800)$ are likely to be sensitive to details of the long range part of the quark-quark interaction because the constituent-quark wave functions will possess material support at large interquark separation.  Hence the development of an understanding of their properties may provide information about light-quark confinement.  The pseudoscalar trajectory is particularly interesting because its lowest mass member is QCD's Goldstone mode.  Therefore an explanation should simultaneously describe: (1) chiral symmetry and its dynamical breaking; and (2) the possible correlation of the trajectory's higher mass members via an approximately linear radial Regge trajectory.  Outcome (2) does not require that confinement in light-quark systems be expressed through the formation of a flux tube.\cite{efimov}  It is easily obtained in Poincar\'e invariant quantum mechanics\cite{klink} but requirement (1) is not.

A Poincar\'e covariant and symmetry preserving treatment of quark-antiquark bound states can be based on the Bethe-Salpeter equation (BSE)
\begin{equation}
\label{bse1}
\Gamma_{tu}(k;P) = \int^\Lambda_q [\chi(q;P)]_{sr}\, K_{rs}^{tu}(q,k;P)\,,
\end{equation}
where: $k$ is the relative and $P$ the total momentum of the constituents; $r$,\ldots,\,$u$ represent colour, Dirac and flavour indices; $\chi(q;P):= S(q_+) \Gamma(q;P) S(q_-)$, $q_\pm = q\pm P/2$; and $\int^\Lambda_q$ represents a translationally invariant regularisation of the four-dimensional integral, with $\Lambda$ the regularisation mass-scale.  In Eq.\,(\ref{bse1}), $S$ is the renormalised dressed-quark propagator and $K$ is the fully amputated dressed-quark-antiquark scattering kernel.  

The dressed-quark propagator appearing in the BSE's kernel is determined by the renormalised gap equation (the quark Dyson-Schwinger equation [DSE])
\begin{eqnarray}
S(p)^{-1} & =&  Z_2 \,(i\gamma\cdot p + m^{\rm bm}) + Z_1 \int^\Lambda_q\! g^2 D_{\mu\nu}(p-q) \frac{\lambda^a}{2}\gamma_\mu S(q) \Gamma^a_\nu(q,p)\,,\label{gendse} 
\end{eqnarray}
where $D_{\mu\nu}$ is the dressed-gluon propagator, $\Gamma_\nu(q,p)$ is the dressed-quark-gluon vertex, $m^{\rm bm}$ is the $\Lambda$-dependent current-quark bare mass, and $Z_{1,2}(\zeta^2,\Lambda^2)$ are, respectively, the quark-gluon-vertex and quark wave function renormalisation constants.   The gap equation's solution has the form 
\begin{eqnarray} 
%\nonumber 
 S(p)^{-1} 
% & = & i \gamma\cdot p \, A(p^2,\zeta^2) + B(p^2,\zeta^2) \,.%\\ 
% 
& =& \left[ i\gamma\cdot p + M(p^2)\right]/Z(p^2,\zeta^2) \,. 
\label{sinvp} 
\end{eqnarray} 
It is obtained from Eq.\,(\ref{gendse}) augmented by the condition $\left.S(p)^{-1}\right|_{p^2=\zeta^2} = i\gamma\cdot p +
m(\zeta)$, where $m(\zeta)$ is the renormalised mass: 
$
Z_2(\zeta^2,\Lambda^2) \, m^{\rm bm}(\Lambda) = Z_4(\zeta^2,\Lambda^2) \, m(\zeta)\,,
$
with $Z_4$ the Lagrangian mass renormalisation constant.  The chiral limit in QCD is unambiguously defined by\cite{langfeld}
$Z_2(\zeta^2,\Lambda^2) \, m^{\rm bm}(\Lambda) \equiv 0 \,, \forall \Lambda \gg \zeta
$,
which is equivalent to stating that the renormalisation-point-invariant current-quark mass $\hat m = 0$.

\begin{figure}[t]
\centerline{\includegraphics[width=0.7\textwidth]{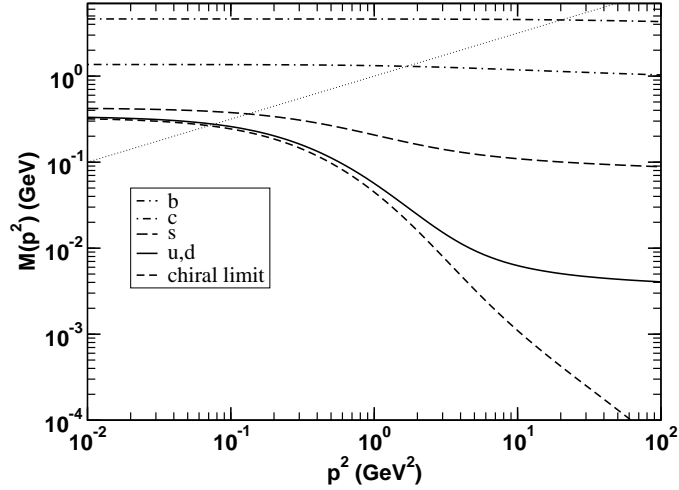}}
\caption{\label{Mp2} Quark mass function obtained with the quenched gap equation kernel described in Ref.~[\protect\refcite{bhagwat}].  At a renormalisation point $\zeta_R=19\,$GeV, the current-quark mass values are: $m_{u,d}(\zeta_R)= 3.7\,$MeV, $m_{s}(\zeta_R)= 82\,$MeV, $m_{c}(\zeta_R)= 0.97\,$GeV, and $m_{b}(\zeta_R)= 4.1\,$GeV.  The indicated solutions of $p^2=M(p^2)^2$ define the Euclidean constituent-quark mass (in GeV): $M_{\hat m = 0}=0.27$, $M_{u,d}=0.28$, $M_{s}=0.36$, $M_{c}=1.3$, and $M_{b}=4.5$.  In perturbation theory $M_{\hat m = 0}(p^2)\neq 0$ is impossible in the chiral limit.  Its appearance here is innate to dynamical chiral symmetry breaking.  The ultraviolet behaviour of the chiral-limit mass function is qualitatively different from that of the solutions obtained with a nonzero current-quark mass.\protect\cite{lane,politzer}  For $p^2\lsim 1\,$GeV$^2$ the $u,d$ mass function is almost indistinguishable from the chiral-limit result: this is the essentially nonperturbative domain.}
\end{figure}

It is a longstanding prediction of DSE studies in QCD that the dressed-quark propagator receives strong momentum-dependent corrections at infrared momenta.\cite{lane,politzer,cdragw}  This phenomenon is illustrated in Fig.\,\ref{Mp2}.  The prediction was confirmed in recent simulations of lattice regularised QCD\cite{bowman} and the conditions under which pointwise agreement may be obtained have been explored.\cite{bhagwat,maris,cdrqcddu,iida}

The pion's properties are fundamentally governed by the phenomenon of dynamical chiral symmetry breaking (DCSB).  One expression of the chiral properties of QCD is the axial-vector Ward-Takahashi identity 
\begin{eqnarray}
P_\mu \Gamma_{5\mu}^j(k;P) & =& S^{-1}(k_+) i \gamma_5\frac{\tau^j}{2}
+  i \gamma_5\frac{\tau^j}{2} S^{-1}(k_-) - \, 2i\,m(\zeta) \,\Gamma_5^j(k;P) ,
\label{avwtim}
\end{eqnarray}
which is here written for two quark flavours, each with the same current-quark mass: $\{\tau^i:i=1,2,3\}$ are flavour Pauli matrices.  In Eq.\,(\ref{avwtim}), $\Gamma_{5\mu}^j(k;P)$ and $\Gamma_5^j(k;P)$ are, respectively, the renormalised dressed-axial-vector and -pseudoscalar vertices, each of which satisfies a unique inhomogeneous extension of Eq.\,(\ref{bse1}).  Equation~(\ref{avwtim}) is an exact statement about chiral symmetry and the pattern in which it is broken.  Hence it must always be satisfied.  Since that cannot be achieved veraciously through fine tuning, the distinct kernels of the DSEs involved; namely, Eq.\,(\ref{gendse}) and the relevant inhomogeneous extensions of Eq.\,(\ref{bse1}), must be intimately related.  Any theoretical tool employed in calculating properties of the pseudoscalar and pseudovector channels must preserve that relationship if the results are to bear fidelity to QCD.  A weak coupling expansion of the DSEs yields perturbation theory and satisfies this constraint.  However, that truncation scheme is not useful in the study of intrinsically nonperturbative phenomena.  Fortunately a nonperturbative systematic and symmetry preserving scheme exists. (Reference\,[\refcite{mandarvertex}] gives details.)  This entails that the full implications of Eq.\,(\ref{avwtim}) can be both elucidated and illustrated.

Since every flavour nonsinglet pseudoscalar meson is exhibited as a pole contribution to the axial-vector and pseudoscalar vertices, it follows from Eq.\,(\ref{avwtim}) that the residues at these poles satisfy\cite{mrt98} \begin{equation}
\label{gmorgen} f_{\pi_n} m_{\pi_n}^2 = 2 \, m(\zeta)  \, 
\rho_{\pi_n}(\zeta)\,, 
\end{equation}
wherein ``$n$'' is the principal quantum number of the $n\, ^1\!S_0$ trajectory with, e.g., $n=0$ corresponding to the ground state $\pi(140)$ pseudoscalar meson, and 
\begin{eqnarray} 
\nonumber f_{\pi_n} \,\delta^{ij} \,  P_\mu = Z_2\,{\rm tr} \int^\Lambda_q \!
\sfrac{1}{2} \tau^i \gamma_5\gamma_\mu\, \chi^j_{\pi_n}(q;P) ,\;
 i  \rho_{\pi_n}\!(\zeta)\, \delta^{ij}  =Z_4\,{\rm tr} 
\int^\Lambda_q \!\sfrac{1}{2} \tau^i \gamma_5 \, \chi^j_{\pi_n}(q;P)\,,\\
\label{fpin}
\end{eqnarray} 
both of which are gauge invariant and cutoff independent.  The identity expressed in Eq.\,(\ref{gmorgen}) is exact in QCD, and valid: for every $0^-$ meson;\cite{krassnigg} and irrespective of the magnitude of the current-quark mass.\cite{mishasvy}

For the ground state meson, DCSB entails $f_{\pi_0}^0 := \lim_{\hat m \to 0}\, f_{\pi_0} \neq 0$ and $\rho_{\pi_0}^0(\zeta) := \lim_{\hat m \to 0}\,\rho(\zeta) = - \langle \bar q q \rangle^0_\zeta/f^0_{\pi_0}\neq 0$.  Hence the Gell-Mann--Oakes--Renner relation is clearly a corollary of Eq.\,(\ref{gmorgen}) and the ground state pion is QCD's Goldstone mode.\cite{mrt98}

For the $n\geq 1$ pseudoscalar mesons: $m_{\pi_{n\geq 1}}>m_{\pi_0}$ by assumption, and hence $m_{\pi_{n> 0}} \neq 0$ in the chiral limit; the existence of a bound state entails that $\chi_n(k;P)$ is a finite matrix-valued function; and, furthermore, the ultraviolet behaviour of the quark-antiquark scattering kernel in QCD guarantees that 
\begin{equation}
\rho_{\pi_{n}}^0(\zeta):= \lim_{\hat m\to 0} \rho_{\pi_{n}}(\zeta) <\infty \,, \; \forall \, n\,.
\end{equation}
Hence, it is a necessary consequence of chiral symmetry and its dynamical breaking in QCD; viz., Eq.\,(\ref{gmorgen}), that\cite{krassnigg}
\begin{equation}
\label{fpinzero}
f_{\pi_n}^0 \equiv 0\,, \forall \, n\geq 1\,.
\end{equation}

These outcomes are legitimate in any theory that has a valid chiral limit.  It is logically possible that such a theory does not exhibit DCSB; i.e., realises chiral symmetry in the Wigner mode.  This is the circumstance, e.g., in QCD above the critical temperature for chiral symmetry restoration.\cite{bastirev}  Equation (\ref{gmorgen}) is still valid in the Wigner phase but its implications are different.\cite{krassnigg}  In this event $f^W_{\pi_0} \propto \hat m$; i.e., the leptonic decay constant of the ground state pseudoscalar meson also vanishes in the chiral limit.  Moreover, since  
$f_{\pi_0}\,\rho_{\pi_{0}}(\zeta) \stackrel{\hat m\approx 0}{\propto} - \langle \bar q q \rangle_\zeta^0$ and $ \langle \bar q q \rangle^{0\,W}_\zeta \propto \hat m^3$ in the Wigner phase,\cite{langfeld} it follows that if a rigorously chirally symmetric theory possesses a massless pseudoscalar bound state then $m^W_{\pi_0} \stackrel{\hat m\approx 0}{\propto} \hat m $.  In this case there is also a degenerate scalar meson partner whose mass behaves in the same manner.

The exact results just reviewed can be illustrated accurately in a model that both preserves QCD's ultraviolet properties and exhibits DCSB.  One such is the renormalisation-group-improved rainbow-ladder DSE model presented in Ref.\,[\refcite{maristandy1}], whose widespread application is reviewed in Ref.\,[\refcite{revpieter}].  The utilisation of this model as an illustrative tool in the present context is described in Ref.\,[\refcite{krassnigg}] and  Table~\ref{tablea} epitomises the calculations.  In regarding the table one should note that the rainbow-ladder truncation gives ``ideal'' flavour mixing and the study used $\hat m_u = \hat m_d \neq \hat m_s \neq \hat m_c$.  Thus the second row simultaneously describes four degenerate pseudoscalar mesons for each $n$, namely, a \{$u \bar d$, $u\bar d-d\bar u$,$d\bar u$\} isotriplet and a $u\bar d+d\bar u$ isosinglet; the third row describes $s\bar s$ mesons; and the fourth, $c\bar c$ mesons.  The isotriplet mesons of Row~2 may plainly be associated with the $n ^1S_0$ pion-like radial trajectory but what should one think of the isosinglet, and what of the $s\bar s$ composites of Row~3?  

\begin{table}[t]
\tbl{\label{tablea} Results for the $0^{-}$ ground and first radially excited states: $m(\zeta_0):= \hat m/(\ln\zeta_0/\Lambda_{\rm QCD})^{\gamma_m}$, $\gamma_m=12/25$, $\zeta_0=1\,$GeV.  As a guide, we list experimental values reported in Ref.\,[\protect\refcite{pdg}] (in GeV): $m_{\pi_0}=0.14\,$, $m_{\pi_1}=1.3\pm 0.1\,$; $m_\eta= 0.548$, $m_{\eta^\prime}= 0.958$, $m_{\eta(2S)}=1.294\pm 0.004$;  $m_{\eta^\prime(2S)}=1.476\pm 0.004$;
$m_{\eta_c(1S)}=2.9796 \pm 0.0012$, $m_{\eta_c(2S)}=3.65\pm 0.01$; 
$f_{\pi_0}=0.092\,$; and\protect\cite{cleo} $f_{\eta_c}= 0.237 \pm 0.053$.  Dimensions are GeV except for $\rho(\zeta)$, GeV$^2$.\vspace*{-4ex}}
{\begin{tabular}{c}
\hspace*{\textwidth}\\
\vspace*{-2ex}
\end{tabular}}
\begin{center}
{\begin{tabular}{ccccccc}
\toprule
\hfill $m(\zeta_0)$ \hfill & \hfill $m_{\pi_0}$ \hfill & \hfill $m_{\pi_1}$ \hfill & \hfill $f_{\pi_0}$ \hfill & \hfill $f_{\pi_1}$ \hfill & \hfill $\rho_{\pi_0}$ \hfill & \hfill $\rho_{\pi_1}$ \hfill \\\colrule
0.0\hphantom{000} & 0.0\hphantom{0} & 1.08 & 0.091 & \hphantom{-}0.0\hphantom{00} & (0.51)$^2$ & -(0.49)$^2$\\
0.0055 & 0.14 & 1.10 & 0.093 & -0.002 & (0.52)$^2$ & -(0.49)$^2$ \\
0.125\hphantom{0} & 0.69 & 1.40 & 0.130 & -0.023 & (0.64)$^2$ & -(0.54)$^2$ \\
1.390\hphantom{0} & 2.98 & 3.42 & 0.233&  -0.109 & (1.15)$^2$ & -(0.84)$^2$\\
\botrule
\end{tabular}}
\end{center}
\end{table}

These questions are answered by a nonzero value of the topological susceptibility in QCD\cite{christos} but this feature is not realised in the rainbow-ladder truncation.  However, starting from the model of Ref.\,[\refcite{mn83}] one may easily explore the resolution of the U(1) problem via a Witten-Veneziano-like mechanism; viz., one describes a rudimentary Bethe-Salpeter equation in which the kernel contains contributions, additional to those of the ladder truncation, whose form is constrained by the properties and consequences of the non-Abelian anomaly.  The isotriplet solutions of this equation are precisely those of the ladder truncation: the non-Abelian anomaly does not alter the fact that the pion is QCD's Goldstone mode.  However, no solution with the ideally-mixed flavour structures $u\bar d+d\bar u$ or $s\bar s$ exists.  Instead, for the ground state pseudoscalar mesons the equation yields two solutions, at masses $m_{\eta_0}$, $m_{\eta_0^\prime}$, with $m_{\pi_0}\leq m_{\eta_0} <m_{\eta_0^\prime}$, whose bound state amplitudes can be represented as: $\eta_0= \cos\theta(\eta_0) {\cal F}^{\bf O} \Gamma^{\bf O}_{\eta_0}(k;P)-  \sin\theta(\eta_0) {\cal F}^{\bf S} \Gamma^{\bf S}_{\eta_0}(k;P)$; and $\eta_0^\prime= \sin\theta(\eta_0^\prime) {\cal F}^{\bf O} \Gamma^{\bf O}_{\eta^\prime_0}(k;P)+ \cos\theta(\eta_0^\prime){\cal F}^{\bf S} \Gamma^{\bf S}_{\eta_0^\prime}(k;P)$, where ${\cal F}^{\bf O}$ represents $(u \bar u + d \bar d - 2 s\bar s)/\surd 6$ and ${\cal F}^{\bf S}$, $(u \bar u + d \bar d +s\bar s)/\surd 3$.  In this way one readily obtains a reasonable picture of the $\eta$-$\eta^\prime$ complex, with $m_{\eta_0} \approx 3.3 m_{\pi_0}$, $m_{\eta_0^\prime} \approx 1.8 \, m_{\eta_0}$ and $\theta(\eta_0)=-10.5^\circ$, $\theta(\eta_0^\prime)=-31^\circ$.  (NB.\, A mass-dependent mixing-angle is natural with the DSEs.\cite{etaenergy} Ideal mixing corresponds to $\theta=-54.7^\circ\!\!$.)  In the chiral limit, $\hat m_{u,d}=\hat m_s = 0$, the model yields: $\theta(\eta_0)=0^\circ = \theta(\eta_0^\prime)$, and hence the $\eta_{0}$ is purely $U(3)$ octet while the $\eta_0^\prime$ is purely $U(3)$ singlet; and $m_{\pi_0}=m_{\eta_0}=0$ while $m^{\hat m = 0}_{\eta_0^\prime}=0.9\, m_{\eta_0^\prime}$.

This model can be adapted to estimate the non-Abelian anomaly's effect on pseudoscalar meson radial excitations.  Thus applied, it predicts $m_{\eta_1} \approx 1.08\,m_{\pi_1} $, $m_{\eta_1^\prime} \approx 1.22 \,m_{\eta_1}$ and $\theta(\eta_1)=-35^\circ$, $\theta(\eta_1^\prime)=-43^\circ$, which matches expectation: the contribution to masses from QCD's nonzero topological susceptibility diminishes with increasing bound state mass.  It follows that the $\eta_1$'s flavour content is $87$\% isosinglet, so that $m_{\eta_1}\gsim m_{\pi_1}$.  Moreover, the $\eta_1^\prime$ is $84$\% $s \bar s$ and hence the rainbow-ladder truncation is again a good approximation.  This discussion suggests that $\eta(1295)$ is the first radial excitation of $\eta(548)$ and, from line~3 of Table~\ref{tablea}, that $\eta(1475)$ might reasonably be interpreted as the first radial excitation of $\eta^\prime(958)$.

The non-Abelian anomaly plays a nugatory role in the spectrum of mesons containing $c$-quarks.  Hence the rainbow-ladder truncation should be realistic.  This may also be inferred from Ref.\,[\refcite{mandarvertex}].  While the light masses were fixed elsewhere,\cite{maristandy1} the $c$-quark mass in Table~\ref{tablea} was chosen to fit $m_{\eta_c(1S)}$.  The values of the other quantities in Row~4 are a prediction.  For model comparison (in GeV): Ref.\,[\refcite{munczekjain}] finds $f_{\eta_c(1S)}=0.213$ and $m_{\eta_c(2S)}=3.47$, and Ref.\,[\refcite{wang}], $f_{\eta_c(1S)}=0.247$.

\begin{figure}[t]
\centerline{\includegraphics[width=0.7\textwidth]{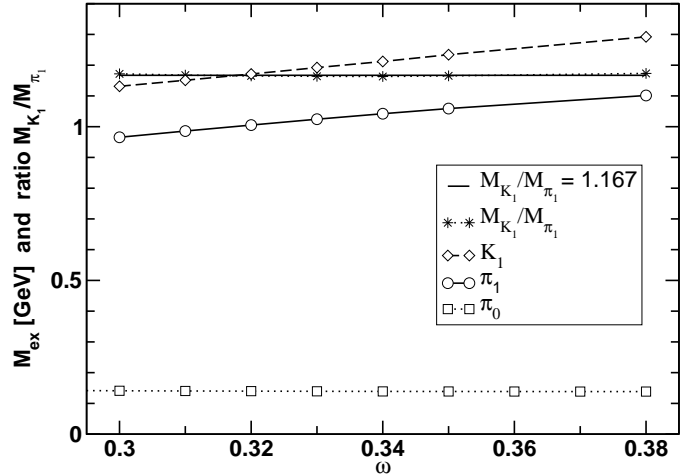}}
\caption{\label{ratio} Evolution of meson masses with $\omega$, where $r_a=1/\omega$ is the range of strong attraction in the model of Ref.\,[\protect\refcite{maristandy1}].  It is apparent that radial excitations are sensitive to the long-range part of the interaction, becoming less bound as the range of strong attraction decreases.}
\end{figure}

The model in Ref.\,[\refcite{maristandy1}] has a single parameter, $\omega$.  It may be used to define a length-scale $r_a = 1/\omega$: magnifying $r_a$ increases the range of strong attraction in the model.  In this connection, Fig.~\ref{ratio} is significant.  It illustrates the evolution of masses with $\omega$ and reveals plainly that the properties of the pion's first radial excitation are sensitive to the long-range part of the interaction; e.g., while a $30$\% increase in $r_a$ raises $m_{\pi_0}$ by merely $3$\%, it reduces $m_{\pi_1}$ by $14$\%.  The sensitivity is explained by the evaluated charge radius:\cite{krassnigg} $r_{\pi_1}=1.7\,r_{\pi_0}$, which indicates that the pion's first radial excitation is a very extended object.  This is materially important for many reasons.  One instance is that a diameter $d_{\pi_1}= 2\,r_{\pi_1}\approx 2.25\,$fm exceeds the spatial length, $L\lsim 2\,$fm, of lattices used in modern numerical simulations of lattice-QCD.

Figure~\ref{ratio} also depicts the $\omega$-evolution of the mass of the kaon's first radial excitation and shows that $m_{K_{n=1}}/m_{\pi_{n=1}} \approx \,$constant.  We use this fact along with the $\pi(2S)$'s experimental mass to obtain $m_{K(2S)}:=m_{K_{n=1}} = 1.52\pm 0.12\,$GeV. This agrees with Ref.\,[\refcite{munczekjain}], and suggests that the state identified as $K(1460)$ is the kaon's first radial excitation.

The Dyson-Schwinger equations (DSEs) provide a unique perspective on nonperturbative aspects of QCD, e.g., they give: a straightforward understanding of the origin of constituent-quark masses; a veracious description of QCD's Goldstone modes; and the essential connection between these phenomena.  The existence of a nonperturbative systematic and symmetry-preserving truncation scheme enables the proof of exact results, which establish that chiral symmetry and the pattern of its breaking have a material impact even on the properties of hadrons with masses above $1\,$GeV.  An appreciation of the essence of this truncation scheme enables the formulation of a phenomenologically efficacious renormalisation-group-improved rainbow-ladder DSE model, corrections to which are quantifiable.  The study of baryons is feasible.\cite{nucleon}  A discussion focusing on QCD's gauge sector may be followed from Ref.\,[\refcite{alkofer}].

\medskip

%\section*{Acknowledgments}
%
\hspace*{-\parindent}{\bf Acknowledgments}.~%
We avouch valuable interactions with P.\ Maris and P.\,C.~Tandy.  CDR thanks the organisers and the staff of IHEP, Beijing, and especially the staff and students in the Department of Physics at Peking University for their peerless hospitality and support.  
This work was also supported by: %
the Austrian Research Foundation \textit{FWF, 
Erwin-Schr\"odinger-Stipendium} no.\ J2233-N08; Department of Energy, 
Office of Nuclear Physics, contract no.\ W-31-109-ENG-38; National Science Foundation contract no.\ INT-0129236; the \textit{A.\,v.\ 
Humboldt-Stiftung} via a \textit{F.\,W.\ Bessel Forschungspreis};  and benefited from the facilities of ANL's Computing Resource Center.

\end{document}